\documentstyle[prb,aps]{revtex}
\begin{document}
\draft
\twocolumn[\hsize\textwidth\columnwidth\hsize\csname @twocolumnfalse\endcsname
\title{Quantum Phonon Optics: \\
Coherent and Squeezed Atomic Displacements}
\author{Xuedong Hu and Franco Nori}
\address
{Department of Physics, The University of Michigan, Ann Arbor, Michigan
48109-1120}
\date{\today}
\maketitle
\begin{abstract}
In this paper we investigate coherent and squeezed quantum states of
phonons.  The latter allow the possibility of modulating the quantum
fluctuations of atomic displacements below the zero-point quantum noise
level of coherent states.  The expectation values and quantum
fluctuations of both the atomic displacement and the lattice amplitude
operators are calculated in these states---in some cases analytically.
We also study the possibility of squeezing quantum noise in the atomic
displacement using a polariton-based approach.
\end{abstract}

\pacs{PACS numbers: 42.50.Lc, 42.50.Dv, 05.40.+j, 63.90.+t}
%
% in the short PRL paper the second number should be the first.
%
\vskip2pc]
\narrowtext

\section{Introduction}

Classical phonon optics \cite{phononopt} has succeeded in producing
many acoustic analogs of {\it classical optics}, such as phonon
mirrors, phonon lenses, phonon filters, and even ``phonon microscopes''
that can generate acoustic pictures with a resolution comparable to
that of visible light microscopy.  Most phonon optics experiments use
heat pulses or superconducting transducers to generate {\em incoherent}
phonons, which propagate ballistically in the crystal.  These ballistic
incoherent phonons can then be manipulated by the above-mentioned
devices, just like in geometric optics.

Phonons can also be excited {\em phase-coherently}.  For instance,
coherent acoustic waves with frequencies of up to $10^{10}$ Hz can be
generated by piezoelectric oscillators.  Lasers have also been used to
generate coherent acoustic and optical phonons through stimulated
Brillouin and Raman scattering experiments.  Furthermore, in recent
years, it has been possible to track the phases of coherent optical
phonons \cite{review}, due to the availability of femtosecond-pulse
ultrafast lasers (with a pulse duration shorter than a phonon period)
\cite{ultra}, and techniques that can measure optical reflectivity with
accuracy of one part in $10^{6}$.

In most situations involving phonons, a {\it classical} description is
adequate.  However, at low enough temperatures, {\it quantum}
fluctuations become dominant.  For example, a recent study \cite{fnori}
shows that quantum fluctuations in the atomic positions can indeed
influence observable quantities (e.g., the Raman line shape) even when
temperatures are not very low.  With these facts in mind, and prompted
by the many exciting developments in {\it classical} phonon optics,
coherent phonon experiments, and (on the other hand) squeezed states of
light \cite{special}, we would like to explore phonon analogs of {\it
quantum} optics. In particular, we study the dynamical and quantum
fluctuation properties of the atomic displacements, in analogy with the
modulation of quantum noise in light.  Specifically, we study
single-mode and two-mode phonon coherent and squeezed states, and then
focus on a polariton-based approach to achieve smaller quantum noise
than the zero-point fluctuations of the atomic lattice.

The concepts of coherent and squeezed states were both originally
proposed in the context of quantum optics.  A coherent state is a
phase-coherent sum of number states.  In it, the quantum fluctuations
in any pair of conjugate variables are at the lower limit of the
Heisenberg uncertainty principle.  In other words, a coherent state is
as ``quiet'' as the vacuum state.  Squeezed states \cite{special} are
interesting because they can have {\it smaller quantum noise than the
vacuum state} in one of the conjugate variables, thus having a
promising future in different applications ranging from gravitational
wave detection to optical communications.  In addition, squeezed states
form an exciting group of states and can provide unique insight
into quantum mechanical fluctuations.  Indeed, squeezed states are now
being explored in a variety of non-quantum-optics systems, including
{\it classical} squeezed states \cite{Rugar}.

In Sec.\ II we introduce some quantities of interest and study the
fluctuation properties of the phonon vacuum and number states.  In
Secs.\ III and IV we investigate phonon coherent and squeezed states.
In Sec.\ V we propose a novel way of squeezing quantum noise in the
atomic displacement operator using a polariton-based mechanism.  The
Appendix summarizes the derivation of the time evolution of the
relevant operators in this polariton approach.  Finally, Sec.\ VI
presents some concluding remarks.

\section{Phonon Operators and the Phonon Vacuum and Number States}

A phonon with quasimomentum ${\bf p} = \hbar {\bf q}$ and branch
subscript $\lambda$ has energy $\epsilon_{{\bf q}\lambda} = \hbar
\omega_{{\bf q} \lambda}$; the corresponding creation and annihilation
operators satisfy the boson commutation relations
\begin{equation}
[b_{{\bf q'}\lambda'},\ b^{\dagger}_{{\bf q}\lambda} ] = \delta_{
{\bf q} {\bf{q'} } }\delta_{\lambda \lambda'} \,, \ \ \ \
[b_{{\bf q}\lambda},\ b_{{\bf q'}\lambda'} ] = 0 \,.
\end{equation}
The atomic displacements $u_{i\alpha}$ of a crystal lattice are given
by
\begin{equation}
u_{i\alpha} = \frac{1}{\sqrt{Nm}} \sum^{N}_{{\bf q} \lambda}
U^{\lambda}_{{\bf q} \alpha}Q^{\lambda}_{\bf q} e^{i {\bf q} \cdot {\bf
R}_i} \,.
\end{equation}
Here ${\bf R}_i$ refers to the equilibrium lattice positions, $\alpha$
to a particular direction, and $Q^{\lambda}_{\bf q}$ is the normal-mode
amplitude operator
\begin{equation}
Q^{\lambda}_{\bf q} = \sqrt{\frac{\hbar}{2 \omega_{{\bf q} \lambda}} }
\left( b_{{\bf q} \lambda} + b^{\dagger}_{-{\bf q} \lambda} \right)
\,.
\end{equation}
An experimentally observable quantity is the real part of the Fourier
transform of the atomic displacement:
\begin{eqnarray}
{\rm Re}\left[ u_{\alpha}({\bf q}) \right] & = & \sum_{\lambda} \sqrt{
\frac{\hbar}{8m \, \omega_{{\bf q}\lambda}} } \left\{ U^{\lambda}_{{\bf
q} \alpha} ( b_{{\bf q}\lambda} + b_{{-\bf q}\lambda}^{\dagger} )
\right.  \nonumber \\
& & \left. + U^{\lambda \; *}_{{\bf q} \alpha} ( b_{{-\bf q}\lambda} +
b_{{\bf q}\lambda}^{\dagger} ) \right\} \,.
\end{eqnarray}
For simplicity, hereafter we will drop the branch subscript $\lambda$,
assume that $U_{{\bf q} \alpha}$ is real, and define a ${\bf q}$-mode
dimensionless lattice amplitude operator:
\begin{equation}
u(\pm {\bf q}) = b_{\bf q} + b^{\dagger}_{-\bf q} + b_{-\bf q} +
b^{\dagger}_{\bf q} \,.
\end{equation}
This operator contains essential information on the lattice dynamics,
including quantum fluctuations.  It is the phonon analog of the
electric field in the photon case.

Let us first consider the phonon vacuum state.  When no phonon is
excited, the crystal lattice is in the phonon vacuum state
$|0\rangle$.  The expectation values of the atomic displacement and the
lattice amplitude are zero, but the fluctuations will be finite:
\begin{eqnarray}
\langle(\Delta u_{i\alpha})^2 \rangle_{\rm vac} & \equiv & \langle
(u_{i\alpha})^2 \rangle_{\rm vac} - \langle u_{i\alpha} \rangle^2_{\rm
vac}  \\
& = & \sum_{\bf q}^{N} \frac{\hbar |U_{{\bf q}
\alpha}|^2}{2Nm\,\omega_{{\bf q}\alpha} } \,, \\
\langle(\Delta u(\pm {\bf q}))^2\rangle_{\rm vac} & = & 2 \,.
\end{eqnarray}

Let us now consider the phonon number states.  The eigenstates of the
harmonic phonon Hamiltonian are number states which satisfy $b_{\bf
q}|n_{\bf q}\rangle = \sqrt{n_{\bf q}} |n_{\bf q}-1\rangle$.  The
phonon number and the phase of atomic vibrations are conjugate
variables.  Thus, due to the uncertainty principle, the phase is
arbitrary when the phonon number is certain, as it is the case with any
number state $|n_{\bf q}\rangle$.  Thus, in a number state, the
expectation values of the atomic displacement $\langle n_{\bf
q}|u_{i\alpha}|n_{\bf q} \rangle$ and ${\bf q}$-mode lattice amplitude
$\langle n_{\bf q}|u(\pm {\bf q})|n_{\bf q} \rangle$ vanish due to the
randomness in the phase of the atomic displacements.  The fluctuations
in a number state $|n_{\bf q}\rangle$ are
\begin{eqnarray}
\langle(\Delta u_{i\alpha})^2\rangle_{\rm num} & = & \frac{\hbar
|U_{{\bf q} \alpha}|^2 n_{\bf q} }{Nm\,\omega_{{\bf q}\alpha} } +
\sum_{{\bf q'} \neq {\bf q}}^{N} \frac{\hbar |U_{{\bf q'}
\alpha}|^2}{2Nm\,\omega_{{\bf q'}\alpha} }  \,, \\
\langle(\Delta u(\pm {\bf q}))^2\rangle_{\rm num} & = & 2 + 2n_{\bf q}
\,.
\end{eqnarray}

\section{Phonon Coherent States}

A single-mode (${\bf q}$) phonon coherent state is an eigenstate of a
phonon annihilation operator:  
\begin{equation}
b_{\bf q}|\beta_{\bf q}\rangle = \beta_{\bf q} |\beta_{\bf q} \rangle
\,.
\end{equation}
It can also be generated by applying a phonon displacement operator
$D_{\bf q}(\beta_{\bf q})$ to the phonon vacuum state
\begin{eqnarray}
|\beta_{\bf q} \rangle & = & D_{\bf q}(\beta_{\bf q}) |0\rangle =
\exp(\beta_{\bf q} b_{\bf q}^{\dagger} - \beta_{\bf q}^* b_{\bf
q})|0\rangle  \\
& = & \exp \left( -\frac{|\beta_{\bf q}|^2}{2} \right) \sum_{n_{\bf
q}=0}^{\infty} \frac{\beta_{\bf q}^{n_{\bf q}}}{\sqrt{n_{\bf q}!}}
|n_{\bf q}\rangle \,.
\end{eqnarray}
Thus it can be seen that a phonon coherent state is a phase-coherent
superposition of number states.  Moreover, coherent states are a set of
minimum-uncertainty states which are as noiseless as the vacuum state.
Coherent states are also the set of quantum states that best describe
the classical harmonic oscillators \cite{Gardiner}.

A single-mode phonon coherent state can be generated by the Hamiltonian
\begin{equation}
H  =  \hbar \omega_{\bf q} \left( b^{\dagger}_{\bf q} b_{\bf q} +
\frac{1}{2} \right) + \lambda^*_{\bf q}(t) \, b_{\bf q} +
\lambda_{\bf q}(t) \, b_{\bf q}^{\dagger}
\end{equation}
and an appropriate initial state.  Here $\lambda_{\bf q}(t)$ represents
the interaction strength between phonons and the external source.  More
specifically, if the initial state is a vacuum state, $|\psi (0)
\rangle = |0 \rangle$, then the state vector becomes a
single-mode coherent state thereafter
\begin{equation}
|\psi(t)\rangle  = |\Lambda_{\bf q}(t) \, e^{-i\omega_{\bf q}t}\rangle
\,,
\end{equation}
where 
\begin{equation}
\Lambda_{\bf q}(t) =  - \frac{i}{\hbar} \int_{-\infty}^t \lambda_{\bf
q}(\tau) \, e^{i\omega_{\bf q}\tau}d\tau
\end{equation}
is the coherent amplitude of mode ${\bf q}$.  If the initial state is a
single-mode coherent state $|\psi(0)\rangle = | \alpha_{\bf q}
\rangle$, then the state vector at time $t$ takes the form
\begin{equation}
|\psi(t)\rangle = |\left\{ \Lambda_{\bf q}(t) + \alpha_{\bf q}
\right\} e^{-i\omega_{{\bf q}}t}\rangle \,,
\end{equation}
which is still a coherent state.

In a single-mode (${\bf q}$) coherent state $|\Lambda_{\bf q}(t) \,
e^{-i\omega_{\bf q}t} \rangle$, $\langle u_{i\alpha}(t) \rangle_{\rm
coh}$ and $\langle u(\pm {\bf q}) \rangle_{\rm coh}$ are sinusoidal
functions of time.  The fluctuation in the atomic displacements is
\begin{equation}
\langle(\Delta u_{i\alpha})^2\rangle_{\rm coh} = \sum_{\bf q}^{N}
\frac{\hbar |U_{{\bf q} \alpha}|^2}{2Nm\,\omega_{{\bf q}\alpha}} \,.
\end{equation}
The unexcited modes are in the vacuum state and thus all contribute to
the noise in the form of zero point fluctuations.  Furthermore,
\begin{equation}
\langle(\Delta u(\pm {\bf q}))^2\rangle_{\rm coh} = 2 \,.
\end{equation}
From the expressions of the noise $\langle (\Delta u_{i \alpha })^2
\rangle_{\rm coh}$ and $\langle (\Delta u[\pm {\bf q})]^2 \rangle_{\rm
coh}$, it is impossible to know which state (if any) has been excited,
while this information is clearly present in the expression of the
expectation value of the lattice amplitude $\langle u(\pm {\bf q})
\rangle_{\rm coh}$.  These results can be straightforwardly generalized
to multi-mode coherent states.

\section{Phonon Squeezed States}

In order to reduce quantum noise to a level below the zero-point
fluctuation level, we need to consider phonon squeezed states.
Quadrature squeezed states are generalized coherent states
\cite{Walls1}.  Here ``quadrature'' refers to the dimensionless
coordinate and momentum.  Compared to coherent states, squeezed ones
can achieve smaller variances for one of the quadratures during certain
time intervals and are therefore helpful for decreasing quantum noise.
Figures~\ref{fig1} and~\ref{fig2} schematically illustrate several
types of phonon states, including vacuum, number, coherent, and
squeezed states.  These figures are the phonon analogs of the
illuminating schematic diagrams used for photons \cite{Walls1}.

A single-mode quadrature phonon squeezed state is generated from a
vacuum state as 
\begin{equation}
|\alpha_{\bf q} , \xi \rangle = D_{\bf q}(\alpha_{\bf
q} ) S_{\bf q}(\xi ) |0\rangle \,;
\end{equation}
a two-mode quadrature phonon squeezed state is generated as
\begin{equation}
|\alpha _{{\bf q}_1}, \alpha _{{\bf q}_2}, \xi \rangle = D_{{\bf q}_1}
(\alpha _{{\bf q}_1})D_{{\bf q}_2} (\alpha _{{\bf q}_2}) S_{{\bf q}_1,
{\bf q}_2} (\xi) |0\rangle \,.
\end{equation}
Here $D_{\bf q}(\alpha_{\bf q})$ is the coherent state displacement
operator with $\alpha_{\bf q} = |\alpha_{\bf q}| e^{i\phi}$, 
\begin{eqnarray}
S_{\bf q} (\xi ) & = & \exp \left( \frac{\xi^*}{2} b_{\bf q}^2 -
\frac{\xi}{2} b_{\bf q}^{\dagger \, 2} \right) \,, \\
S_{{\bf q}_1, {\bf q}_2} (\xi ) & = & \exp \left( \xi^* b_{{\bf q}_1}
b_{{\bf q}_2} - \xi b_{{\bf q}_1}^{\dagger} b_{{\bf q}_2}^{\dagger}
\right) \,,
\end{eqnarray}
are the single- and two-mode squeezing operator, and $\xi=r
e^{i\theta}$ is the complex squeezing factor with $r \geq 0$ and $ 0
\leq \theta < 2\pi$.  The squeezing operator $S_{{\bf q}_1, {\bf
q}_2}(\xi)$ can be produced by the following Hamiltonian:
\begin{eqnarray}
H_{{\bf q}_1, {\bf q}_2} & = & \hbar \omega_{{\bf q}_1}
b^{\dagger}_{{\bf q}_1} b_{{\bf q}_1} + \hbar \omega_{{\bf q}_2}
b_{{\bf q}_2}^{\dagger} b_{{\bf q}_2}  \nonumber \\
& & + \zeta(t) b^{\dagger}_{{\bf q}_1} b^{\dagger}_{{\bf q}_2} +
\zeta^*(t)  b_{{\bf q}_1} b_{{\bf q}_2} \,.
\end{eqnarray}
The time-evolution operator has the form
\begin{equation}
U(t) = \exp \left( - \frac{i}{\hbar} H_0 t \right) \exp \left[ \xi^*(t)
b_{{\bf q}_1}b_{{\bf q}_2} - \xi(t) b_{{\bf q}_1}^{\dagger} b_{{\bf
q}_2}^{\dagger} \right] \,,
\end{equation}
where 
\begin{equation}
H_0 =  \hbar \omega_{{\bf q}_1} b^{\dagger}_{{\bf q}_1} b_{{\bf q}_1} +
\hbar \omega_{{\bf q}_2} b^{\dagger}_{{\bf q}_2} b_{{\bf q}_2} \,,
\end{equation}
and
\begin{equation}
\xi(t) =  \frac{i}{\hbar} \int_{-\infty}^t \zeta(\tau) \,
e^{i(\omega_{{\bf q}_1} + \omega_{{\bf q}_2}) \tau} d\tau \,.
\end{equation}
Here
$\xi(t)$ is the squeezing factor and $\zeta(t)$ is the strength of the
interaction between the phonon system and the external source; this
interaction allows the generation and absorption of two phonons at a
time.  The two-mode phonon quadrature operators have the form 
\begin{eqnarray}
X({\bf q},{-\bf q}) & = & 2^{-3/2}\left( b_{\bf q} + b_{\bf
q}^{\dagger} + b_{-\bf q} + b_{-\bf q}^{\dagger} \right)   \\
& = & 2^{-3/2}u(\pm {\bf q}) \,, \\
P({\bf q},{-\bf q}) & = & -i 2^{-3/2} \left( b_{\bf q} - b_{\bf
q}^{\dagger} + b_{-\bf q} - b_{-\bf q}^{\dagger} \right)  \,.
\end{eqnarray}

We have considered two cases where squeezed states were involved in
modes $\pm {\bf q}$.  In the first case, the system is in a two-mode
($\pm {\bf q}$) squeezed state $|\alpha_{\bf q}, \alpha_{-\bf q}, \xi
\rangle$, ($\xi = re^{i\theta}$), and its fluctuation is
\begin{equation}
\langle [\Delta u(\pm {\bf q})]^2 \rangle = 2 \left( e^{-2r} \cos^2
\frac{\theta}{2} + e^{2r} \sin^2 \frac{\theta}{2} \right) \,.
\end{equation}
In the second case, the system is in a single-mode squeezed state
$|\alpha_{\bf q}, \xi \rangle$ ($\alpha_{\bf q} = |\alpha_{\bf
q}|e^{i\phi}$) in the first mode and an arbitrary coherent state
$|\beta_{-\bf q} \rangle$ in the second mode.  The fluctuation is now
\begin{eqnarray}
\langle [\Delta u(\pm {\bf q})]^2 \rangle & = & 1 + e^{2r} \sin^2
\left(\phi + \frac{\theta}{2} \right)  \nonumber \\
& & + e^{-2r}\cos^2 \left(\phi + \frac{\theta}{2} \right) \,.
\end{eqnarray}
In both of these cases, $\langle [\Delta u(\pm {\bf q})]^2 \rangle$ can
be smaller than in coherent states (see Fig.~2).

\section{Polariton Approach}

Phonon squeezed states can be generated through phonon-phonon
interactions.  This will be discussed elsewhere \cite{xhu}.  Here we
focus on how to squeeze quantum noise in the atomic displacements
through phonon-photon interactions.  When an ionic crystal is
illuminated by light, there can be a strong coupling between photons
and the local polarization of the crystal in the form of phonons.
Photons and phonons with the same wave vector can thus form polaritons
\cite{Madelung}.  Although now phonons and photons are not separable in
a polariton, we can still study the quantum noise in the atomic
displacements.  Let us consider the simplest Hamiltonian
\cite{Madelung} describing the above scenario:  
\begin{eqnarray}
H_{\rm polariton} & = & \sum_{\bf k} \left\{ E_{1{\bf k}}\,
a^{\dagger}_{\bf k}a_{\bf k} + E_{2{\bf k}} b^{\dagger}_{\bf k}\ b_{\bf
k} \right.  \nonumber \\
& & \hspace{-0.3in} \left. + E_{3{\bf k}}\, \left( a^{\dagger}_{\bf k}
b_{\bf k} - a_{\bf k} b^{\dagger}_{\bf k} - a_{\bf k} b_{-{\bf k}} +
a^{\dagger}_{-{\bf k}} b^{\dagger}_{\bf k} \right) \right\} \,,
\end{eqnarray}
where 
\begin{eqnarray}
E_{1{\bf k}} & = & \hbar c k \,, \\
E_{2{\bf k}} & = & \hbar \omega_0 \sqrt{1 + \chi} \,, \\
E_{3{\bf k}} & = & i\left( \frac{\hbar^2 ck\omega_0 \chi}{4\sqrt{1 +
\chi}} \right)^{1/2} \,.
\end{eqnarray}
Here ${\bf k}$ is the wave vector for both photons and phonons and
$\omega_0$ is the bare phonon frequency.  $\chi$ is the dimensionless
dielectric susceptibility of the crystal (the strength of the
phonon-photon interaction) defined by
\begin{equation}
\chi \omega_0^2
\varepsilon_0 {\bf E} = \ddot{\bf P} + \omega_0^2{\bf P} \,,
\end{equation}
where ${\bf E}$ is the electric field of the incoming light and ${\bf
P}$ is the polarization generated by optical phonons in the crystal.
In $H_{\rm polariton}$, the two free oscillator sums correspond to free
photons and free phonons, while the mixing terms come from the
interaction ${\bf E} \cdot {\bf P}$ between photons and phonons.  The
phonon energy $E_{2{\bf k}}$ has been corrected as $\omega_0$ is
substituted by $\omega_0 \sqrt{1 + \chi}$, so that we have ``dressed''
phonons.

Our goal is to compute the fluctuations of the lattice amplitude
operator $u(\pm {\bf k},t) = b_{\bf k}(t) + b^{\dagger}_{-\bf k}(t) +
b_{-\bf k}(t) + b^{\dagger}_{\bf k}(t)$.  In a two-mode ($\pm {\bf k}$)
coherent state $|\alpha_{\bf q},\,\alpha_{-\bf q}\rangle$, its variance
is $\langle [\Delta u(\pm {\bf q})]^2 \rangle_{\rm coh} = 2$.
Therefore, if at any given time we obtain a value less than $2$, the
lattice amplitude of the relevant mode is squeezed.  In our
calculation, we diagonalize the polariton Hamiltonian and find the
time-dependence of $u(\pm {\bf q})$.  The Appendix presents in more
detail the derivation of the time-evolution of $u(\pm {\bf q})$.

Our results show that the fluctuation property of $u(\pm {\bf q})$
sensitively depends on the $t = 0$ initial state $|\psi(0) \rangle$ of
both phonons and photons.  Our results are summarized in
Table~\ref{table:1}, and some numerical examples are shown in
Fig.~\ref{fig3}.  These calculations focus on the case where $ck$ is
close to $\omega_0$ (the bare phonon frequency, which is typically
$\sim 10$ THz for optical phonons) and thus our typical time is $\sim
0.1$ ps.  More specifically, squeezing effects in $u(\pm {\bf
k})$ are relatively strong for either one of the following two sets of
$t = 0$ initial states:  $(i)$ photon and phonon coherent states, or
$(ii)$ single-mode photon squeezed state and phonon vacuum state.  For
instance, the maximum squeezing exponent $r$ is $0.015$ when the
incident photon state has a squeezing factor $\xi = 0.1 e^{2ickt}$
(where $ck$ is the photon frequency).  On the other hand, with an
initial two-mode photon ($\pm {\bf k}$) squeezed state and two-mode
($\pm {\bf k}$) phonon vacuum state, the squeezing effect in $u(\pm
{\bf k})$ is weak.  We have also used initial conditions with a
single-mode photon squeezed state and thermal states in the two phonon
modes.

Figure~\ref{fig4} shows the temperature dependence of the squeezing
effect for several values of the dielectric susceptibility $\chi$ of
the crystal.  Our numerical results show that squeezing effects are
quickly overshadowed by the thermal noise for small $\chi$, while for
large $\chi$ (e.g., $\chi = 0.5$) the squeezing effect can exist up to
$T \approx 250$ K, as illustrated in Fig.~\ref{fig4}.

\section{Conclusions}

In conclusion, we have investigated the dynamics and quantum
fluctuation properties of phonon coherent and squeezed states.  In
particular, we calculate the experimentally observable time evolution
and fluctuation of the lattice amplitude operator $u(\pm {\bf q})$.  We
show that the $\langle u(\pm {\bf q}) \rangle$ are sinusoidal functions
of time in both coherent and squeezed states, but the fluctuation
$\langle (\Delta u(\pm {\bf q}))^2 \rangle$ in a squeezed phonon state
is periodically smaller than its vacuum or coherent state value $2$.
Therefore, phonon squeezed states are periodically quieter than the
vacuum state.  In the polariton approach to squeezing, we calculate the
atomic displacement part of a polariton, and prove that the
fluctuations of the associated lattice amplitude operator can be
squeezed for different combinations of initial photon and phonon states
and large enough ($\chi > 0.1$) interaction strength.

It is difficult to generate squeezed states because they have noise
levels which are even lower than the one for the vacuum state.  Indeed,
the experimental and theoretical development of photon coherent and
squeezed states took decades.  Likewise, the experimental realization
of phonon squeezed states might require years of further theoretical
and experimental work.  Nevertheless, we believe that theoretical
results in quantum phonon optics can help the development of the
corresponding experiments.  We hope that our effort \cite{xhu} into
this very rich problem will lead to more theoretical and experimental
developments in the still unexplored area of quantum phonon optics and
the manipulation of phonon quantum fluctuations.

\section*{Acknowledgments}

It is a great pleasure for us to acknowledge useful conversations with
Saad Hebboul, Roberto Merlin, Nicolas Bonadeo, Hailin Wang, Duncan
Steel, Jeff Siegel, and especially Shin-Ichiro Tamura.

\appendix

\section{Time-evolution of the Phonon Operators in a Polariton}

To derive the time-evolution of the phonon operators, we need to first
diagonalize the polariton Hamiltonian $H_{\rm polariton}$.  For this
purpose, we introduce the polariton operators $\alpha_{i{\bf k}}$ in
terms of the phonon and photon operators $b_{\bf k}$ and $a_{\bf k}$
\begin{equation}
\alpha _{i{\bf k}} = w_i a_{\bf k} + x_i\ b_{\bf k} + y_i \ a^{\dagger}
_{-{\bf k}} + z_i\ b^{\dagger}_{-{\bf k}} \,, \ \ \ \ i = 1,2 \,.
\end{equation}
If we write
\begin{eqnarray}
{\bf \alpha} & = & (\alpha_{1{\bf k}}, \,\alpha_{2{\bf k}},
\,\alpha_{1,-{\bf k}}^{\dagger}, \,\alpha_{2,-{\bf k}}^{\dagger})^{T}
\,, \\
{\bf a} & = & (a_{\bf k}, \,b_{\bf k}, \,a_{-{\bf k}}^{\dagger},
\,b_{-{\bf k}}^{\dagger})^{T} \,,
\end{eqnarray}
the above relation can be written in a matrix form
\begin{equation}
{\bf \alpha} = {\bf A} \cdot {\bf a}  \,,
\end{equation}
with its inverse ${\bf a} = {\bf A}^{-1} \cdot {\bf \alpha}$.  Here
${\bf A}$ is a matrix given by
\begin{equation}
{\bf A} = \left( 
	\begin{array}{cccc} 
	  w_1 & x_1 & y_1 & z_1 \\
	  w_2 & x_2 & y_2 & z_2 \\ 
	  y_1^* & z_1^* & w_1^* & x_1^* \\
	  y_2^* & z_2^* & w_2^* & x_2^*  
	\end{array} \right) \,.
\end{equation}

In the polariton representation, the Hamiltonian has the diagonal form
\begin{eqnarray}
H_{\rm polariton}^{\prime} & = & \sum_{\bf k} \ \left[ E^{(1)}_{\bf k}
\left(\alpha ^{\dagger}_{1{\bf k}} \ \alpha _{1{\bf k}} +
\frac{1}{2}\right)\right.  \\
& & \left. + E^{(2)}_{\bf k}\ \left(\alpha ^{\dagger}_{2{\bf
k}}\ \alpha _{2{\bf k}} + \frac{1}{2}\right) \right] \, .
\end{eqnarray}
The subindices $i = 1\,,2\,$ specify the two polariton branches, with
different dispersion relations $E^{(1)}_{\bf k}$ and $E^{(2)}_{\bf
k}$.  The transformation matrix elements $w_i,\ x_i,\ y_i,\ z_i$ are
determined by requiring that the $\alpha _{i{\bf k}}\,$'s satisfy boson
commutation relations
\begin{equation}
[\alpha_{i{\bf k}} \, , \ \alpha_{j{\bf k}'}^{\dagger}] = \delta_{ij}
\, \delta_{{\bf k}{\bf k}'}\,,  \ \ \ [\alpha_{i{\bf k}} \,,
\ \alpha_{j{\bf k}'}] = 0 \,,
\end{equation}
so that
\begin{equation}
[\alpha_{i{\bf k}} \; , \ H] = E^{(i)}_{\bf k}\ \alpha _{i{\bf k}}\,,
%\label{eq:commutator}
\end{equation}
which is true if the two different polariton branches are independent
of each other.  

In the polariton representation, the Hamiltonian $H_{\rm
polariton}^{\prime}$ describes two independent harmonic oscillators.
From the Heisenberg equation
\begin{equation}
i\,\hbar\frac{d\hat O}{ dt} = [\hat O ,\ H] \,,
\end{equation}
we obtain 
\begin{mathletters}
\begin{eqnarray}
\alpha_{1{\bf k}}(t) & = & \alpha_{1{\bf k}}(0) \ e^{-i
E^{(1)}_{\bf k} \, t / \hbar } \,,  \\
\alpha_{2{\bf k}}(t) & = & \alpha_{2{\bf k}}(0) \ e^{-i
E^{(2)}_{\bf k} \, t / \hbar } \,,
\end{eqnarray}
\end{mathletters}
or in a more compact form
\begin{equation}
{\bf \alpha}(t) = {\bf U}_{\alpha}(t) \, {\bf \alpha}(0) \,.
\end{equation}
Recall that the matrix form of the canonical transformation from the
photon and phonon operators ($a_{\bf k}$ and $b_{\bf k}$) to the
polariton operators ($\alpha_{\bf k}$) is ${\bf \alpha} = {\bf A} \cdot
{\bf a}$.  Thus, at time $t$ the photon and phonon operators can be
expressed as
\begin{equation}
{\bf a}(t) = {\bf A}^{-1} \, {\bf \alpha}(t) = {\bf A}^{-1} \, {\bf
U}_{\alpha}(t) \, {\bf A} \, {\bf a}(0) \,,
\end{equation}
which provides the time evolution of the photon and phonon operators.

%\vspace{-0.1in}

%\newpage
\begin{table}[h]
%\begin{center}
\begin{tabular}{|l|c|l|}
	$t = 0$ photons & $t = 0$ phonons & 
	Squeezed $u(\pm {\bf k})$?  \\  \hline \hline
	CS($\pm {\bf k}$) & CS($\pm {\bf k}$) & yes (no) if $\chi >
	(\leq) \ 0.1$ \\ \hline
	SMST(${\bf k}$), VS(${-\bf k}$) & VS($\pm {\bf k}$) & yes (no)
	if $\chi > (\leq) \ 0.1$  \\ \hline
	SMST(${\bf k}$), VS(${-\bf k}$) & TS($\pm {\bf k}$) & yes if
	$T < T_s(\chi)$ \\ \hline
	TMST($\pm {\bf k}$) & VS($\pm {\bf k}$) & weak (no) if $\chi >
	(\leq) \ 0.1$ \\
\end{tabular}
\caption{Different combinations of $t = 0$ initial states (modes $\pm
{\bf k}$) for the polariton approach to lattice amplitude squeezing
and the corresponding effects in the fluctuations of the lattice
amplitude operator $u(\pm {\bf k})$.  Here CS(${\bf k}$), VS(${\bf
k}$), TS(${\bf k}$), SMST(${\bf k}$),  TMST($\pm {\bf k}$) refer,
respectively, to coherent, vacuum, thermal, single-mode and two-mode
squeezed states in the mode inside the parentheses, ${\bf k}$ or $\pm
{\bf k}$.  $T_s(\chi)$ is the temperature below which squeezing is
obtained (see Fig.~4).  By squeezing we mean that the quantum noise of
the relevant variable is below its corresponding vacuum state value.}
%\end{center}
\label{table:1}
\end{table}

%\newpage

\begin{figure}[h]
\caption[]{Schematic diagram of the uncertainty areas
(shaded) in the generalized coordinate and momentum ($X({\bf q},{-\bf
q})$, $\,P({\bf q},{-\bf q})$) phase space of (a) the phonon vacuum
state, (b) a phonon number state, (c) a phonon coherent state, and (d)
a phonon squeezed state.  Here $X({\bf q},{-\bf q})$ and $P({\bf
q},{-\bf q})$ are the two-mode ($\pm {\bf q}$) coordinate and momentum
operators, which are the direct generalizations of their corresponding
single-mode operators.  Notice that the phonon coherent state has the
same uncertainty area as the vacuum state, and that both areas are
circular, while the squeezed state has an elliptical uncertainty area.
Therefore, in the direction parallel to the $\theta/2$ line, the
squeezed state has a smaller noise than both the vacuum and coherent
states.}
\label{fig1} 
\end{figure}

\begin{figure}[h]
\caption[]{Schematic diagram of the time evolution of the expectation
value and the fluctuation of the lattice amplitude operator $u(\pm {\bf
q})$ in different states.  Dashed lines represent $\langle u(\pm {\bf
q}) \rangle$, while the solid lines represent the envelopes $\langle
u(\pm {\bf q}) \rangle \pm \sqrt{\langle [\Delta u(\pm {\bf q})]^2
\rangle}$.  (a) The phonon vacuum state $|0\rangle$, where $\langle
u(\pm {\bf q}) \rangle = 0$ and $\langle [\Delta u(\pm {\bf q})]^2
\rangle = 2$.  (b) A phonon number state $|n_{\bf q}, n_{-\bf
q}\rangle$, where $\langle u(\pm {\bf q}) \rangle = 0$ and $\langle
[\Delta u(\pm {\bf q})]^2 \rangle = 2(n_{\bf q} + n_{-\bf q}) + 2$.
(c) A single-mode phonon coherent state $|\alpha_{\bf q} \rangle$,
where $\langle u(\pm {\bf q}) \rangle = 2|\alpha_{\bf q}| \cos
{\omega_{\bf q}t}$ (i.e., $\alpha_{\bf q}$ is real), and $\langle
(\Delta u[\pm {\bf q})]^2 \rangle = 2$.  (d) A single-mode phonon
squeezed state $|\alpha_{\bf q}e^{-i\omega_{\bf q}t}, \, \xi(t)
\rangle$, with the squeezing factor $\xi(t) = re^{-2i\omega_{\bf q}t}$
and $r = 1$.  Here, $\langle u(\pm {\bf q}) \rangle = 2|\alpha_{\bf
q}| \cos {\omega_{\bf q}t}$, and $\langle [\Delta u(\pm {\bf q})]^2
\rangle = 2(e^{-2r} \cos^2 {\omega_{\bf q}t} + e^{2r}\sin^2{\omega_{\bf
q}t})$.  (e) A single-mode phonon squeezed state, as in (d); now the
expectation value of $u$ is $\langle u(\pm {\bf q}) \rangle =
2|\alpha_{\bf q}| \sin {\omega_{\bf q}t}$, (i.e.  $\alpha_{\bf q}$ is
purely imaginary), and the fluctuation $\langle [\Delta u(\pm {\bf
q})]^2 \rangle$ has the same time-dependence as in (d).  Notice that
the squeezing effect now appears at the times when the lattice
amplitude $\langle u(\pm {\bf q}) \rangle$ reaches its maxima, while in
(d) the squeezing effect is present at the times when $\langle u(\pm
{\bf q}) \rangle$ is close to zero.}
\label{fig2} 
\end{figure}

\begin{figure}
\caption[]{Calculated $\langle [\Delta u(\pm {\bf k})]^2
\rangle$ versus time for different combinations of photon and phonon
initial states using a polariton mechanism for lattice amplitude
squeezing.  Dashed (solid) lines correspond to a susceptibility $\chi =
0.1 \; (0.4)$.  Time is measured in units of $1/ck$, where $ck$ is the
free photon frequency.  These calculations focus on the case where $ck$
is close to $\omega_0$ (the bare phonon frequency, which is typically
$\sim 10$ THz for optical phonons) and thus our typical time is $\sim
0.1$ ps.  The horizontal lines at $\langle [\Delta u(\pm {\bf
k})]^2 \rangle = 2$ correspond to the noise level of coherent states.
Thus, any time the fluctuation satisfies $\langle [\Delta u(\pm {\bf
k})]^2 \rangle < 2$ (highlighted), the state is squeezed.  Different
combinations of initial states were considered.  (a) Photon and phonon
coherent states.  (b) Single-mode squeezed state in photon mode ${\bf
k}$ with squeezing factor $\xi = 0.1$ and a vacuum state in the photon
mode ${-\bf k}$; both phonon modes are in the vacuum state.  (c) Same
combination of states as in (b), but here $\xi = 0.1 e^{2it}$.}
\label{fig3}
\end{figure}

\begin{figure}
\caption[]{Temperature dependence of the {\it minimum} fluctuation
min$\{\langle (\Delta u[\pm {\bf k})]^2 \rangle\}$ in $u(\pm {\bf k})$
using a polariton mechanism for squeezing.  The phonon frequency is
$10$ THz.  The initial states are:  single-mode squeezed state in
photon mode ${\bf k}$, vacuum state in photon mode $-{\bf k}$, and
thermal state in both ($\pm {\bf k}$) phonon modes.  The squeezing
factor is $\xi = 0.1 e^{2it}$.  Squeezing can exist up to a temperature
$T_s(\chi)$.  For example, when $\chi = 0.2$, squeezing effects vanish
when $T \agt 25$ K.  On the other hand, for stronger photon-phonon
interaction (e.g., $\chi = 0.5$), the squeezing effects can be obtained
up to $T \approx 250$ K.}
\label{fig4}
\end{figure}

\end{document}